\documentclass[aps,prl,superscriptaddress,preprint]{revtex4-2}
\usepackage{amsfonts}
\usepackage{xcolor}
\usepackage{graphicx}
\usepackage{mathtools}
\usepackage{etoolbox}
\usepackage{setspace}
\makeatletter
\patchcmd{\frontmatter@abstract@produce}
  {\vskip200\p@\@plus1fil
   \penalty-200\relax
   \vskip-200\p@\@plus-1fil}
  {}
  {}
  {}
\makeatother

\allowdisplaybreaks

\renewcommand{\eqref}[1]{Eq.~(\ref{#1})}

\begin{document}

\title{Topological pumping of bimerons in spiral magnets}

\author{Luca Maranzana}
\affiliation{Quantum Materials Theory, Italian Institute of Technology, Via Morego 30, Genoa, Italy}
\affiliation{Department of Physics, University of Genoa, Via Dodecaneso 33, Genoa, Italy}

\author{Maxim Mostovoy}
\affiliation{Zernike Institute for Advanced Materials, University of Groningen, Nijenborgh 3, 9747 AG Groningen, Netherlands}

\author{Naoto Nagaosa}
\affiliation{RIKEN Center for Emergent Matter Science (CEMS), Wako, Saitama 351-0198, Japan}
\affiliation{Fundamental Quantum Science Program (FQSP), TRIP Headquarters, RIKEN, Wako 351-0198, Japan}

\author{Sergey Artyukhin}
\affiliation{Quantum Materials Theory, 16162, Genoa, Italy}

\begin{abstract}
Precise positioning of topological defects is essential for racetrack memories, where their positions along a magnetic nanotrack encode information. Traditional methods achieve nanometric precision by engineering pinning landscapes that enforce discrete steps in defect motion. However, accessing each bit requires overcoming a depinning threshold, which increases power consumption. Here, we demonstrate that spiral magnets provide a natural ruler, enabling precise positioning of bimerons---topological spin textures analogous to skyrmions---without relying on engineered pinning sites. A rotating magnetic field couples directly to the bimeron position, displacing it by exactly one spiral period per full rotation of the field. Such quantized transport of skyrmionic textures, reminiscent of Thouless pumping, is topologically protected and remains robust against perturbations, positioning spiral magnets as a natural skyrmion racetrack. The findings establish a paradigm for topologically protected transport of spin textures.
\end{abstract}

\maketitle

\section{Introduction}
Localized topological defects in magnetic materials have attracted enormous interest and recently gave birth to the field of skyrmionics \cite{Skyrme62, Polyakov75, Bogdanov94, Nagaosa13}. Magnetic skyrmions are nanometer-sized topologically stable spin textures that hold promise for new efficient information storage and processing devices, such as skyrmion racetrack memories \cite{Parkin08, Fert13}.
While skyrmions are usually realized on a ferromagnetic background, spiral multiferroics provide the simplest noncollinear background that breaks translation and inversion symmetries (see Fig.~\ref{fig:Bimeron}(a)) \cite{Kimura03, Katsura05, Mostovoy06}, leading to improper ferroelectricity, nonlocal domain wall dynamics and pinning \cite{Li12, Foggetti22, Maranzana24}. Here, we show that spiral multiferroics host bimerons, skyrmion-like localized topological defects consisting of vortex-antivortex pairs (see Fig.~\ref{fig:Bimeron}(b, c)) \cite{Gross78, Kharkov17, Gobel19, Kim19, Zhang20, Pozzi21, Bachmann23}. These spin textures carry both magnetic and ferroelectric dipole moments, allowing their manipulation through electric and magnetic fields. The dynamics of bimerons on a spiral background is a problem of both fundamental and technological importance, as the interaction between the bimeron and the spiral results in novel mechanisms for the manipulation of skyrmionic textures.

In racetrack memories, information is encoded in the spatial arrangement of domain walls or skyrmionic textures \cite{Parkin08, Fert13}, which are shifted along a track to align with the read and write heads. Therefore, robust position control is essential for storing and processing information. Existing approaches rely on engineered pinning landscapes, created by implanting impurities or modulating the device topography. However, these methods increase power consumption, as each bit must overcome a depinning threshold to be accessed. Here, we demonstrate that bimerons can be precisely positioned on the spiral background by rotating a magnetic field, eliminating the need to engineer pinning sites.
Unlike in a collinear background, where the field drives the bimeron only indirectly through excitation of high-frequency modes \cite{Wang15, Moon16, DelSer23, Shen20}, the spiral background enables direct control over its position. Specifically, a slowly rotating field deforms the bimeron energy landscape, shifting its minima along the spiral. As a result, the bimeron is adiabatically pumped by exactly one period of the spiral for each full rotation of the field. Remarkably, this mechanism does not require the spiral itself to move---unlike an Archimedean screw \cite{DelSer21, Kurebayashi22}---and is therefore unaffected by spiral pinning \cite{Foggetti22}. Furthermore, the nontrivial topology of the pump cycle ensures robustness against perturbations, akin to Thouless pumping \cite{Thouless83}. Depending on the magnetic coupling between adjacent spiral chains, bimerons can move either along or perpendicular to the spiral wave vector. The longitudinal motion is entirely driven by topological pumping, while the transverse component arises from Berry-phase contributions. Both adiabatic and non-adiabatic pumping regimes are analyzed, with the transition occurring at a critical rotation frequency set by the field amplitude. The findings, built upon the symmetries of the spiral background, are applicable to a broad class of spiral magnets, including spiral multiferroics, and are expected to extend to more complex topological spin textures, such as hopfions \cite{Faddeev97, Sutcliffe17, Voinescu20, Azhar24}.

\section{Bimerons on a spiral background}
Consider the following Ginzburg-Landau free energy for a centrosymmetric spiral magnet in two dimensions,
\begin{equation}
\label{eq:H0}
    H_0 = \int{dxdy \left(J_x \left[-Q^2 (\partial_x {\bf m})^2 + \frac{1}{2} (\partial_{x}^2 {\bf m})^2\right] + \frac{1}{2} \lvert J_\perp \rvert (\partial_y {\bf m})^2 + K_z m_z^2\right)},
\end{equation}
where ${\bf m}(x,y)$ is a unit vector indicating the local magnetization direction, and the lengths (such as $x$, $y$, and $Q^{-1}$) are expressed in units of the lattice constant.
The term proportional to $J_x > 0$ induces a rotation of ${\bf m}(x,y)$ in space with the wave vector $\pm Q \hat{x}$. $J_\perp < 0$ describes a ferromagnetic interaction along $y$. We treat the case of antiferromagnetic $J_\perp$ in a dedicated section. $K_z > 0$ corresponds to an easy $xy$-plane anisotropy.
Therefore, the minimum energy configurations are two spirals where ${\bf m}$ rotates in the $xy$-plane with chirality $\chi = \pm 1$,
\begin{equation}
\label{eq:Spiral}
    {\bf m}_\chi(x) = \left(\cos(Qx - \phi), \chi \sin(Qx - \phi), 0\right).
\end{equation}
The phase $\phi$ describes a translation of the entire spiral along $x$. In what follows, we consider the spiral in which spins rotate counterclockwise as we move in the $x$-direction, i.e. $\chi = +1$.

Two-dimensional ferromagnets host metastable states classified by the topological charge (skyrmion number) \cite{Skyrme62, Polyakov75}
\begin{equation}
\label{eq:TopQ}
    q_\mathrm{top} = \frac{1}{4\pi} \int{dxdy \: {\bf m} \cdot \left[\partial_x {\bf m} \times \partial_y {\bf m}\right]}\,.
\end{equation}
For an easy-axis anisotropy, the topological defect with $q_\mathrm{top} = 1$ is a skyrmion \cite{Skyrme62, Polyakov75, Bogdanov94}. Instead, for an easy-plane anisotropy, it is a bimeron, namely a vortex-antivortex pair with opposite $m_z$ at the centers of the two defects \cite{Gross78, Kharkov17, Gobel19, Kim19, Zhang20, Pozzi21, Bachmann23}. Remarkably, spiral states minimizing \eqref{eq:H0}, as well as conical spiral states \cite{Pozzi21}, also host bimerons.
\begin{figure*}[b]
\includegraphics[width=0.99\linewidth]{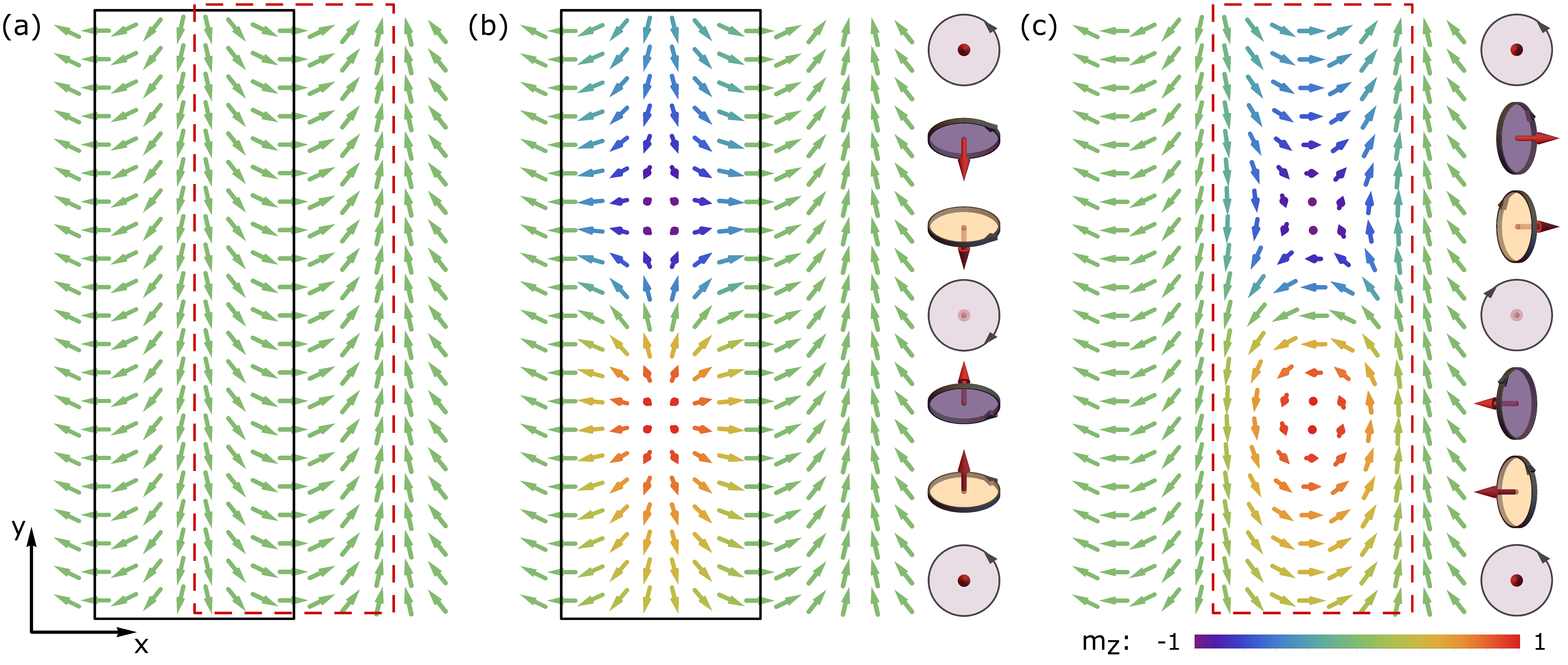}
    \caption{\label{fig:Bimeron}(a) The spiral background breaks translational and rotational symmetries. Black and red boxes indicate the two stripes of width $\Delta x = \pi/Q$ at different positions relative to the spiral where the bimeron is inserted in panels (b) and (c). The colors encode the magnetization component $m_z$. The spiral plane rotates by $2\pi$ about $\hat{x}$, within the black box (b), and about $\hat{y}$, within the red box (c), upon moving along $-y$. These twists of the spiral plane, shown in the right insets of panels (b) and (c), correspond to bimerons with vortex helicities $\varphi = 0$ and $\varphi = \pi/2$, respectively. For either case, the spin texture consists of a vortex with positive $m_z$ (red) and an antivortex with negative $m_z$ (blue), each with topological charge $1/2$. Therefore, the total topological charge of the bimeron is $1$, equal to that of a skyrmion.}
\end{figure*}
As illustrated in Fig.~\ref{fig:Bimeron}(b), a bimeron can be constructed by rotating the spins within the black box about the $x$-axis. The rotation angle increases from $0$ at the top of the box to $2\pi$ at the bottom. Thus, the bimeron twists the spiral plane by $2\pi$ in the $y$-direction. To ensure continuity of the magnetization texture at the boundary of the box, its width along $x$ must be $\pi/Q$, and the spin rotation axis must be parallel to the spins at the vertical boundaries. Hence, this axis changes to the $y$-axis as we move the box and, therefore, the bimeron by $\pi/(2Q)$ along the $x$-direction (see Fig.~\ref{fig:Bimeron}(c)). In general, the spin rotation axis is $\hat{e}_\varphi = \left(\cos{\varphi},\sin{\varphi},0\right)$, where $\varphi = Q\bar{x} - \phi + \pi/2$ depends on the bimeron $x$-coordinate, $\bar{x}$. We observe that $\varphi$ corresponds to the vortex helicity, namely the angle between the in-plane spin component and the radius vector from the vortex center. We encode the previous considerations into an Ansatz for the spin texture of a bimeron with position $(\bar{x},\bar{y})$:
\begin{equation}
\label{eq:Ansatz}
    {\bf m}(x,y) = \left\{
    \begin{aligned}
    &\hat{\mathcal{R}}_{\hat{e}_\varphi}(\theta(y-\bar{y}))\,{\bf m}_+(x)& &\qquad -\frac{\pi}{2 Q} < x - \bar{x} < \frac{\pi}{2 Q} \\
    &{\bf m}_+(x)& &\qquad \mathrm{otherwise}
    \end{aligned}
    \right.,
\end{equation}
where ${\bf m}_+(x) \equiv {\bf m}_{\chi\,=\,+1}(x)$ denotes the counterclockwise spiral \eqref{eq:Spiral} and $\hat{\mathcal{R}}_{\hat{e}_\varphi}(\theta(y-\bar{y}))$ is a rotation matrix around $\hat{e}_\varphi$ by an angle
\begin{equation}
\label{eq:Theta}
    \theta(y-\bar{y}) = 2 q_\mathrm{top} \arccos{\!\left(\!\tanh{\!\left(\frac{y-\bar{y}}{\Tilde{\lambda}}\right)}\!\right)}.
\end{equation}
Minimizing \eqref{eq:H0} with respect to $\Tilde{\lambda}$, we obtain $\Tilde{\lambda} = \sqrt{3\lvert J_\perp \rvert / 2K_z} \equiv \sqrt{3/2}\,\lambda$. 
The $2\pi$ twist of the spiral plane described by $\hat{\mathcal{R}}_{\hat{e}_\varphi}(\theta(y-\bar{y}))$ endows the bimeron with an integer topological charge $q_\mathrm{top} = \pm1$ and ensures its stability. The latter property is verified by relaxing the spin configuration \eqref{eq:Ansatz} via the LLG dynamics \cite{Landau35, Gilbert04, Skubic08} (see Appendix~B). We observe that the Ansatz \eqref{eq:Ansatz} presents a derivative discontinuity in $\partial_x{\bf m}(x,y)$. In the actual magnetization texture, this discontinuity is smoothed out by a spiral distortion near the vertical boundaries of the boxes in Fig.~\ref{fig:Bimeron}. Such a deformation costs an energy $\delta E \approx J_x Q^4 \delta x \lambda$, where $\delta x\lesssim Q^{-1}$ is the width of the distortion. Since this energy is proportional to $\lambda$, exactly as the anisotropy energy $E_{K_z} \approx K_z Q^{-1} \lambda$, the bimeron length scale along $y$ is given by $\lambda \approx \sqrt{\lvert J_\perp \rvert / (K_z + K_Q)}$, with $K_Q \propto J_x Q^4$. In general, the denominator of $\lambda$ collects all the energy contributions that scale linearly with the bimeron size in the $y$-direction.

To avoid the limitations of a specific Ansatz, we proceed with an Ansatz-free derivation of the general bimeron properties from its symmetries and length scales. The bimeron texture is invariant under the following two transformations of coordinates and spins:
\begin{eqnarray}
    x-\bar{x} \rightarrow -(x-\bar{x}) &\quad\mathrm{and}\quad& (m_x,m_y) \rightarrow \hat{\mathcal{R}}_{\hat{z}}(2\varphi)(-m_x,m_y), \label{eq:Sym1} \\
    y-\bar{y} \rightarrow -(y-\bar{y}) &\quad\mathrm{and}\quad& m_z \rightarrow -m_z, \label{eq:Sym2}
\end{eqnarray}
where $\hat{\mathcal{R}}_{\hat{z}}(2\varphi)$ is a rotation matrix in the $xy$-plane by an angle $2\varphi$, double the helicity angle. 


Bimerons carry ferroelectric and magnetic dipole moments
\begin{eqnarray}
    d^{(E)}_z &=& \gamma \int{dxdy \: \left( \hat{x}\cdot[{\bf m}\times\partial_y{\bf m}] - \hat{y}\cdot[{\bf m}\times\partial_x{\bf m}] \right)} = C_E \frac{\gamma q_\mathrm{top}}{Q} \sin(Q\bar{x} - \phi),\label{eq:MomE} \\
    {\bf d}^{(M)} &=& \mu \int{dxdy \: {\bf m}} = -C_M \frac{\mu\lambda}{Q} \left(\cos(Q\bar{x} - \phi), \sin(Q\bar{x} - \phi), 0\right) \propto - {\bf m}_+(\bar{x}),\label{eq:MomH}
\end{eqnarray}
where $\gamma$ is the magnetoelectric coupling constant \cite{Mostovoy06}, $\mu$ is the magnetic moment of a single spin, the bimeron length scales in the $x$ and $y$ directions are $Q^{-1}$ and $\lambda$, and we imposed the constraint on $\varphi$ discussed above. Using the Ansatz \eqref{eq:Ansatz}, we can estimate the dimensionless shape factors $C_E$ and $C_M$: $C_E = \pi^2$ and $C_M = 4\sqrt{6}$. The components of ${\bf d}^{(M)}$ perpendicular to ${\bf m}_+(\bar{x})$ are zero due to \eqref{eq:Sym1} and \eqref{eq:Sym2}. Similarly, $\hat{y}\cdot[{\bf m}\times\partial_x{\bf m}]$ does not contribute to the integral, because of \eqref{eq:Sym2}. Remarkably, the dipole moments depend on the bimeron position $\bar{x}$. In particular, ${\bf d}^{(M)}$ is opposite to the magnetization of the spiral background at the bimeron center, ${\bf m}_+(\bar{x})$. Bimerons also carry an electric dipole moment in the $y$-direction $d^{(E)}_y = \gamma \int{dxdy \: \left(\hat{z}\cdot[{\bf m}\times\partial_x{\bf m}] - \hat{z}\cdot[{\bf m}_+\times\partial_x{\bf m}_+]\right)}$, as spins near the bimeron center rotate with the opposite chirality with respect to the spiral background ${\bf m}_+(x)$. Hence, the bimeron constitutes the smallest ferroelectric domain in spiral multiferroics. However, $d^{(E)}_y$ does not affect the dynamics described in this work since it is independent of the bimeron position.

Using \eqref{eq:MomE} and \eqref{eq:MomH}, the bimeron energy in the presence of an electric field $E_z$ and a magnetic field ${\bf H}$ takes the form, at the first order in the applied fields,
\begin{equation}
\label{eq:H}
    H = H_0 - C_E \frac{\gamma q_\mathrm{top}}{Q} E_z \sin(Q\Delta\bar{x}) + C_M \frac{\mu\lambda}{Q}\, {\bf H} \cdot \left(\cos(Q\Delta\bar{x}), \sin(Q\Delta\bar{x}), 0\right) \equiv H_0 + U_{\Delta\bar{x}},
\end{equation}
where $\Delta\bar{x} = \bar{x} - \phi/Q$ is the bimeron position relative to the spiral background. The applied fields favor a specific position for the bimeron,
modulo the spiral period $2 \pi/Q$.
This result is closely connected to the screw symmetry of the background: the spiral in \eqref{eq:Spiral} is invariant under the combined action of a rotation of ${\bf m}$ in the $xy$-plane by an angle $\phi$ and a translation along $x$ by $\phi/Q$. Hence, $\Delta\bar{x}$ must involve both translation and rotation to move the bimeron along $x$ without altering the spiral background. At zero applied fields, the energy \eqref{eq:H0} is invariant under these two transformations. Consequently, the bimeron can freely move with respect to the spiral background. However, the applied fields break isotropy in the $xy$-plane and, therefore, the symmetry associated with $\Delta\bar{x}$. The bimeron now experiences a potential $U_{\Delta\bar{x}}$ that depends on its position relative to the spiral $\Delta\bar{x}$. In-plane anisotropy also breaks rotational symmetry and contributes to $U_{\Delta\bar{x}}$. For simplicity, we first focus on the effects of the fields, postponing the discussion of in-plane anisotropy to a later section.

\section{Topological pumping of bimerons}
\begin{figure}[b]
\includegraphics[width=0.99\linewidth]{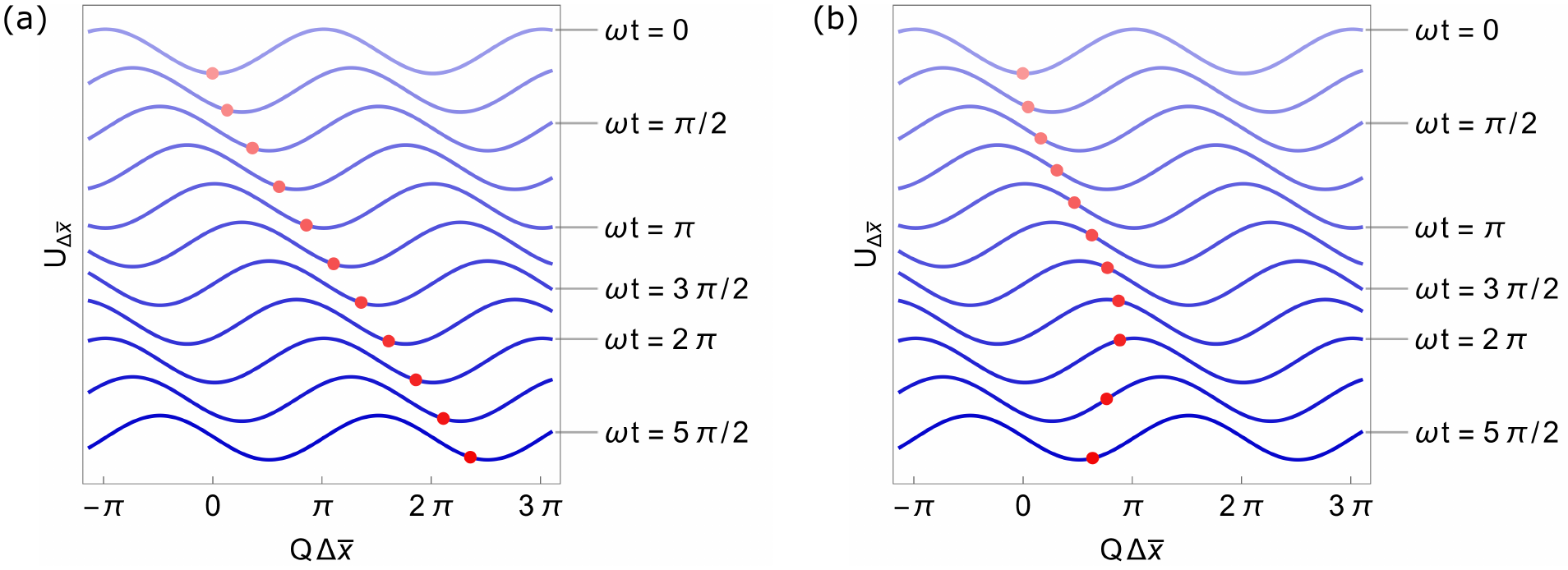}
    \caption{\label{fig:Potential}Effective potential induced by the rotating magnetic field $U_{\Delta\bar{x}}$ as a function of the bimeron position relative to the spiral background $\Delta\bar{x}$, plotted at different times $t$. (a) For a slowly rotating field, the bimeron (red dot) follows the sliding potential $U_{\Delta\bar{x}}$, residing near a minimum. (b) Above a critical rotation frequency, the bimeron is no longer confined within a well of $U_{\Delta\bar{x}}$ because of the gyrotropic and damping forces.}
\end{figure}

Consider the rotating magnetic field ${\bf H}(t) = \lvert{\bf H}\rvert \left(\cos(\omega t),\sin(\omega t),0\right)$. The resulting term in the bimeron energy \eqref{eq:H} takes the form $U_{\Delta\bar{x}} \propto \cos(Q\Delta\bar{x} - \omega t)$, a sinusoidal potential that moves with velocity $\omega/Q$. For a slowly rotating field, the bimeron adiabatically follows one of the minima of the potential $\Delta\bar{x}_\mathrm{min}^{(n)} = (\omega/Q)t + (2n + 1)\pi/Q$ with $n \in \mathbb{Z}$ and therefore moves along $x$ by one spiral period for each full rotation of the magnetic field (see Fig.~\ref{fig:Potential}(a)), resembling the Archimedean screw. Unlike in typical realizations of this mechanism \cite{DelSer21, Kurebayashi22}, the spiral background remains fixed while the bimeron is pumped relative to it. In general, a time-periodic deformation of the effective potential $U_{\Delta\bar{x}}$, such as that induced by a rotating magnetic field, defines a closed trajectory in the parameter space of the system. After a full cycle, $U_{\Delta\bar{x}}$ returns to its original form. However, since the potential is also periodic in space with the spiral period, this does not imply that the positions of its minima remain the same. They may shift by an integer multiple $w$ of the spiral period. Formally, $w$ can be defined as the winding number of the map that associates the phase of the rotating field (an angle on the circle $\mathcal{S}^1$) with the position of a minimum modulo the spiral period (also an element of $\mathcal{S}^1$). Provided the bimeron adiabatically follows a sliding minimum, the integer $w$ classifies the topology of the pumping cycle: a nonzero value signals quantized transport by $w$ spiral periods per rotation of the field. For the rotating-field protocol introduced above, $w=1$, in direct analogy with Thouless pumping \cite{Thouless83}.

To describe the motion along $y$ and the non-adiabatic regime of this bimeron topological pump, we consider the following Lagrangian and Rayleigh dissipation functional that encode the dynamics of a magnetization texture ${\bf m}(t,{\bf r})$ \cite{Landau35, Gilbert04, Duine10},
\begin{equation}
\label{eq:LeR}
    L = \int{dxdy \: {\bf A}({\bf m}) \cdot \dot{{\bf m}}} - H, \qquad\qquad R = \frac{\alpha}{2} \int{dxdy \: \dot{{\bf m}}^{\!2}},
\end{equation}
where the Berry connection ${\bf A}({\bf m})$ satisfies $\nabla_{\!\bf m} \times {\bf A}({\bf m}) = {\bf m}$, the time is redefined according to $t' = t/S$ with $S$ denoting the spin, and $\alpha \ll 1$ is the Gilbert damping constant.


The dynamics described by \eqref{eq:LeR} can be solved exactly only in the simplest cases \cite{Schryer74}. Thus, we apply the collective coordinates approach \cite{Tretiakov08, Clarke08}. In this framework, the dynamics of a magnetization texture is formulated in terms of the collective coordinates $\xi_I(t)$, so that ${\bf m}(t,{\bf r}) = {\bf m}(\{\xi_I(t)\},{\bf r})$. Although ${\bf m}(t,{\bf r})$ possess infinitely many $\xi_I(t)$, only the \textit{soft} modes with low characteristic frequencies, compared to the timescales of the dynamics, are relevant. The other \textit{hard} modes adiabatically adjust to their equilibrium values.
Here, we explore the dynamics induced by a magnetic field that rotates significantly slower than the characteristic frequencies determined by \eqref{eq:H0}, e.g. $\omega_\mathrm{res} = \sqrt{10 K_z J_x Q^4}$ that corresponds to the resonant sliding of the spiral background \cite{DelSer21}. Hence, only the modes associated with the continuous symmetries of \eqref{eq:H0}---$\Delta\bar{x}(t)$, $\bar{y}(t)$ and $\phi(t)$---can be excited.
The spiral phase $\phi(t)$ and the bimeron $y$-coordinate $\bar{y}(t)$ correspond to global translations along $x$ and $y$: $\partial_\phi = -(1/Q)\partial_x$ and $\partial_{\bar{y}} = -\partial_y$. The bimeron $x$-coordinate relative to the spiral background $\Delta\bar{x}(t)$ moves the bimeron along $x$ without altering the spiral, as it involves the screw symmetry discussed at the end of the previous section.

Substituting ${\bf m}(t,{\bf r}) = {\bf m}(\{\xi_I(t)\},{\bf r})$ in \eqref{eq:LeR}, one obtains an effective Lagrangian and Rayleigh dissipation functional, where the dynamical variables are only the soft modes $\xi_I(t)$. The corresponding equations of motion take the form \cite{Tretiakov08, Clarke08}
\begin{equation}
\label{eq:Thiele}
    \alpha \Gamma_{IJ} \dot{\xi}_J - G_{IJ} \dot{\xi}_J = F_I,
\end{equation}
where $I$ and $J$ run over the soft modes, $\Delta\bar{x}(t)$, $\bar{y}(t)$ and $\phi(t)$.
The damping matrix $\Gamma_{IJ} = \Gamma_{JI}$, the gyrotropic matrix $G_{IJ} = - G_{JI}$ and the conservative forces $F_I$ acting on $\xi_I$ descend from $R$, ${\bf A}({\bf m}) \cdot \dot{{\bf m}}$ and $H$, respectively. In Appendix~A, we derive these terms for a bimeron under the action of a rotating magnetic field.

\begin{figure}[b]
\includegraphics[width=0.99\linewidth]{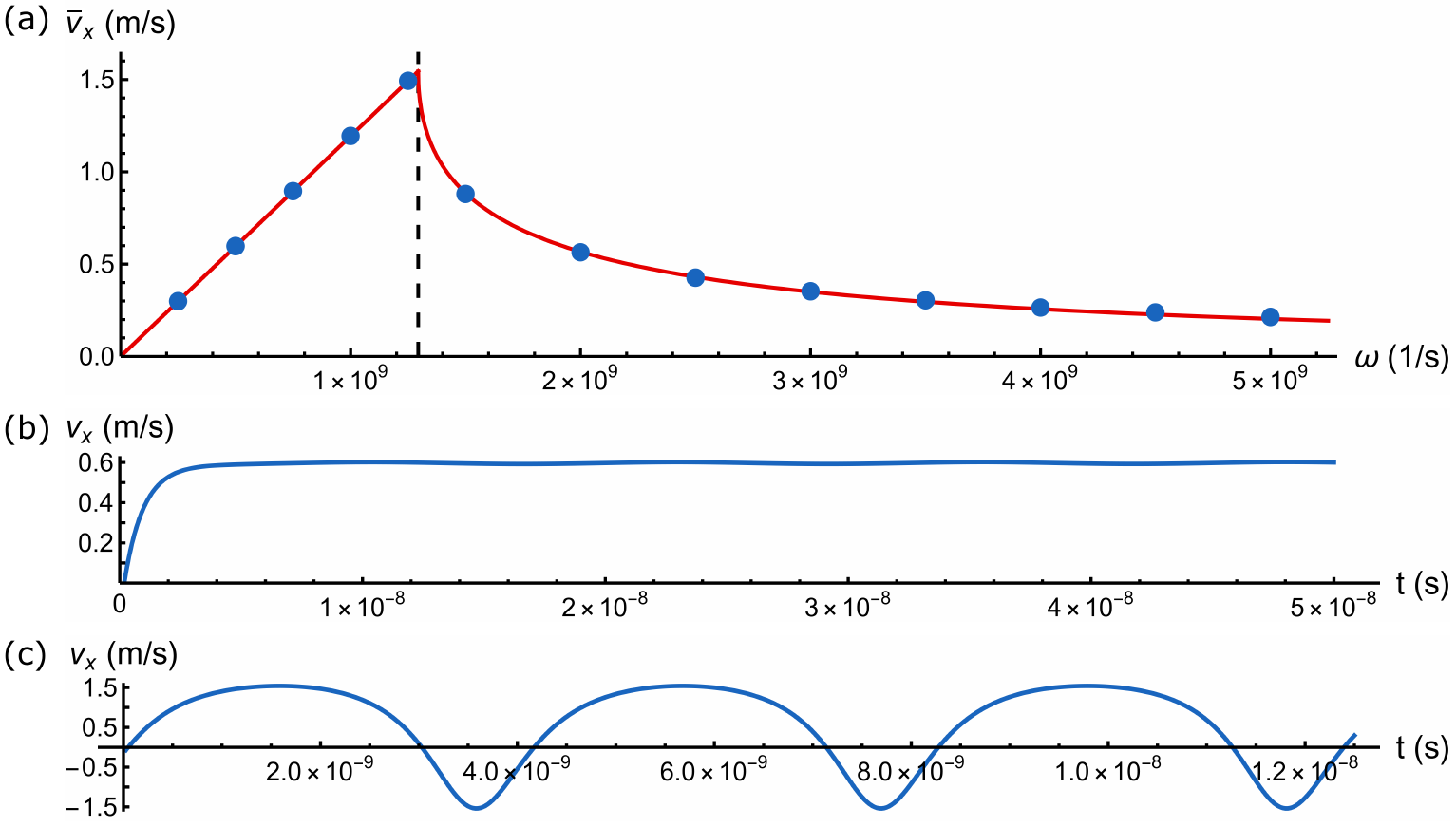}
    \caption{\label{fig:Breakdown}(a) Time-averaged velocity  $\bar{v}_x$ of a bimeron moving under the action of a rotating magnetic field with angular frequency $\omega$. Blue dots indicate the numerical result of the LLG equation \cite{Landau35, Gilbert04, Skubic08} for $Q = 8.38\cdot10^8\,\mathrm{m}^{-1}$, $\lvert{\bf H}\rvert = 0.1\;\mathrm{T}$, $\alpha = 0.1$ and $\mu = \mu_\mathrm{B}$, where $\mu_\mathrm{B}$ is the Bohr magneton. The red line shows the analytical solution \eqref{eq:2Reg} with fitted critical angular frequency $\omega^* = 1.29\cdot10^9\,\mathrm{s}^{-1}$. Using \eqref{eq:EqSteady}, we estimate $\omega^*$ analytically: $\omega^* = 1.12\cdot10^9\,\mathrm{s}^{-1}$. Panels (b) and (c) show the bimeron velocity $v_x = \Delta\dot{\bar{x}}$ as a function of time in the two dynamical regimes. (b) For $\omega = 5\cdot10^8\,\mathrm{s}^{-1} < \omega^*$, $v_x $ reaches the steady-state velocity after a transient in which the bimeron adjusts to its equilibrium position on the sliding potential shown in Fig.~\ref{fig:Potential}(a). (c) For $\omega = 2\cdot10^9\,\mathrm{s}^{-1} > \omega^*$, $v_x$ oscillates with a nonzero average $\bar{v}_x$, as the bimeron no longer follows a minimum of the potential (see Fig.~\ref{fig:Potential}(b)).}
\end{figure}

For large systems, with size $L_x L_y \gg \lambda/(Q\alpha^2)$, the velocity of the spiral $v_\mathrm{s} = \dot{\phi}/Q$ becomes
\begin{equation}
\label{eq:Phi}
v_\mathrm{s} = -\frac{16 \pi^2}{C_{\bar{y}}}\frac{\lambda }{Q\alpha^2  L_x L_y} \!\:\Delta\dot{\bar{x}}.
\end{equation}
Using the Ansatz \eqref{eq:Ansatz}, we can estimate the dimensionless shape factor $C_{\bar{y}}$: $C_{\bar{y}} = 4\pi\sqrt{2/3}$. As the bimeron moves relative to the spiral with a velocity $\Delta\dot{\bar{x}}$, the entire spiral background moves in the opposite direction.
Since the spiral experiences a damping proportional to the system size, its velocity $v_\mathrm{s} \propto (L_x L_y)^{-1}$ is negligible with respect to the bimeron velocity $\Delta\dot{\bar{x}}$. We observe that the magnetic field induces a spiral distortion that propagates along $x$ as ${\bf H}$ rotates. The amplitude of this distortion is resonantly enhanced when the rotation frequency $\omega$ coincides with $\omega_\mathrm{res} = \sqrt{10 K_z J_x Q^4}$ \cite{DelSer21}. This effect contributes to the spiral velocity with a term independent of the system size, which is present even in the absence of the bimeron. However, for $\omega \ll \omega_\mathrm{res}$, the spiral distortion is a hard mode and its contribution is negligible. Within this approximation, the other two equations resulting from \eqref{eq:Thiele} take the form
\begin{equation}
\label{eq:EqFields2}
    \Delta\dot{\bar{x}} = \frac{\alpha\mu C_{\bar{y}}C_M}{16\pi^2 Q} \lvert{\bf H}\rvert \sin(Q\Delta\bar{x} - \omega t), \qquad\qquad \dot{\bar{y}} = -\frac{4 \pi q_\mathrm{top} Q\lambda}{\alpha C_{\bar{y}}} \!\:\Delta\dot{\bar{x}}.
\end{equation}
The bimeron moves along $y$, perpendicular to the spiral wave vector, with a velocity $\dot{\bar{y}}$ that is proportional to the velocity along $x$, $\Delta\dot{\bar{x}}$, and to the topological charge, $q_\mathrm{top}$. This result can be explained considering each term in \eqref{eq:Thiele} as a force. The gyrotropic term, proportional to $G_{IJ}$, results in the ``Lorentz'' force ${\bf F}_G \propto q_\mathrm{top} {\bf v}_\mathrm{b} \times \hat{z}$, where ${\bf v}_\mathrm{b}$ is the bimeron velocity. As it moves in the $x$-direction, following the sliding potential due to the rotating magnetic field $U_{\Delta\bar{x}}$, the bimeron is pushed in the $y$-direction by ${\bf F}_G$. The resulting $\dot{\bar{y}}$ leads to a $x$-component of ${\bf F}_G$, which drives the bimeron away from the minimum of $U_{\Delta\bar{x}}$. This effect combines with the damping force ${\bf F}_\Gamma \propto -\alpha {\bf v}_\mathrm{b}$ and, above a critical ${\bf v}_\mathrm{b}$, overcomes the barrier that confines the bimeron in one of the wells of the sliding sinusoidal potential $U_{\Delta\bar{x}}$ (see Fig.~\ref{fig:Potential}(b)). Thus, the topological pump enters a new dynamical regime in which the bimeron no longer follows $U_{\Delta\bar{x}}$.
To describe such a transition, we first seek a solution of the form $\Delta\bar{x} = (\omega/Q)t + \Delta x_0$, where the bimeron follows the sliding potential. The first of \eqref{eq:EqFields2} becomes
\begin{equation}
\label{eq:EqSteady}
    1 \geq \sin(Q\Delta x_0) = \frac{16\pi^2\omega}{\alpha\mu C_{\bar{y}}C_M \lvert{\bf H}\rvert} \quad\Longrightarrow\quad \omega \leq \frac{\alpha\mu C_{\bar{y}}C_M \lvert{\bf H}\rvert}{16\pi^2} \equiv \omega^*.
\end{equation}
This steady-state solution exists only if the rotation frequency of the magnetic field is below a critical threshold $\omega^*$, which corresponds to the critical bimeron velocity $\Delta\dot{\bar{x}}^* = \omega^*/Q$. In such a case, the position of the bimeron relative to the sliding potential $U_{\Delta\bar{x}}$ takes the form $\Delta x_0 = \left((2n + 1)\pi - \arcsin(\omega/\omega^*)\right)/Q$ with $n \in \mathbb{Z}$. Therefore, the bimeron is displaced from a minimum of the potential by $d = - \arcsin(\omega/\omega^*)/Q$, so that the conservative force exerted by $U_{\Delta\bar{x}}$ compensates for the gyrotropic and damping forces.
At $\omega = \omega^*$, the displacement $d$ corresponds to the steepest point of $U_{\Delta\bar{x}}$, and the conservative force is the maximum possible. Hence, at higher frequencies, the sliding potential can no longer confine the bimeron and no stationary solution exists. The time-averaged bimeron velocity along the $x$-direction $\bar{v}_x$ still takes a simple form, and we have (see Fig.~\ref{fig:Breakdown})
\begin{equation}
\label{eq:2Reg}
    \bar{v}_x = \Delta\dot{\bar{x}} = \frac{\omega}{Q} \quad \mathrm{for}\ \omega \leq \omega^*, \qquad\qquad \bar{v}_x = \frac{\omega}{Q} \left( 1 - \sqrt{1-\frac{\omega^{*2}}{\omega^2}} \right) \!\quad \mathrm{for}\ \omega > \omega^*.
\end{equation}
To summarize, the rotating magnetic field with angular frequency $\omega$ induces a translation of the effective potential for the bimeron with a velocity $\omega/Q$. For $\omega \leq \omega^*$, the bimeron follows the sliding potential. Consequently, its velocity increases linearly with $\omega$ and is remarkably independent of the field strength and the Gilbert damping constant.
This adiabatic pumping shifts the bimeron by one spiral period per field rotation, while keeping the spiral background fixed. The corresponding winding number is $w=1$, reflecting the topological nature of the cycle. At a critical angular frequency $\omega^*$, proportional to the strength of the rotating field, such an adiabatic pumping breaks down due to the gyrotropic and damping forces, and the dynamics sharply enters a new regime where the bimeron velocity decreases as $\omega$ increases. The bimeron also moves perpendicular to the spiral wave vector with a velocity $\dot{\bar{y}}$ inversely proportional to the Gilbert damping constant $\alpha \ll 1$ (see \eqref{eq:EqFields2}). Therefore, $\dot{\bar{y}}$ dominates over $\Delta\dot{\bar{x}}$, in contrast to the antiferromagnetic case treated in a later section.
All the results are expressed for a counterclockwise spin rotation in the spiral background \eqref{eq:Spiral}, $\chi = +1$. In the clockwise case, $\chi = -1$, the velocity of the bimeron is opposite in sign. We corroborate the analytical solution with spin dynamics simulations \cite{Skubic08} (see Fig.~\ref{fig:Breakdown} and Appendix~B).

Domain walls in easy-axis ferromagnets exhibit a behavior analogous to \eqref{eq:2Reg} under a rotating magnetic field in the hard-plane \cite{Yan09, Kim20}. The domain wall position $x_\mathrm{DW}$ and helicity $\varphi_\mathrm{DW}$ play the roles of the bimeron coordinates $\bar{y}$ and $\Delta\bar{x} = (\varphi - \pi/2)/Q$, respectively. Since the ferromagnetic background is translationally invariant, topological pumping can only act on a purely rotational mode, $\varphi_\mathrm{DW}$. Therefore, in contrast to the bimeron case, no quantized domain wall transport can occur.

\section{In-plane anisotropy and topology of the pumping cycle}
We now study the effect of in-plane anisotropy $K_y$ on the bimeron adiabatic pumping that occurs for $\omega \leq \omega^*$. As discussed in the previous section, the pumping cycle is characterized by an integer winding number $w$, which is equal to 1 for $K_y = 0$. In the presence of in-plane anisotropy, the following term must be added to the energy of a bimeron, \eqref{eq:H}:
\begin{equation}
\label{eq:HKy}
    H_{K_y} = \int{dxdy \: K_y m_y^2} = -C_{K_y} K_y \frac{\lambda}{Q}\sin^2{\!(Q\Delta\bar{x})}.
\end{equation}
With the Ansatz \eqref{eq:Ansatz}, the shape factor becomes $C_{K_y} = 2\pi\sqrt{2/3}$. Unlike the contribution of the rotating magnetic field, the static potential in \eqref{eq:HKy} has half the spiral period. For $\lvert{\bf H}\rvert < \mathrm{H}^* = C_{K_y} \lvert K_y \rvert /  C_M \mu$, the maximum force due to the field $F_\mathrm{max}^{({\bf H})} = C_M \mu\lambda \lvert{\bf H}\rvert$ is smaller than the one exerted by the anisotropy $F_\mathrm{max}^{(K_y)} = C_{K_y} K_y \lambda$. Hence, the bimeron is confined in one of the minima of the static potential in \eqref{eq:HKy}, and the pumping cycle is topologically trivial ($w=0$). Conversely, for $\lvert{\bf H}\rvert > \mathrm{H}^*$, the field overcomes the anisotropy and shifts the minima of the effective potential by one spiral period per field rotation. Thus, the winding number is $w=1$, and the pumping cycle is topologically equivalent to the $K_y = 0$ case. This demonstrates that the adiabatic pumping of bimerons is robust against in-plane anisotropy provided the field exceeds $\mathrm{H}^*$.

\begin{figure}[b]
	\includegraphics[width=0.99\linewidth]{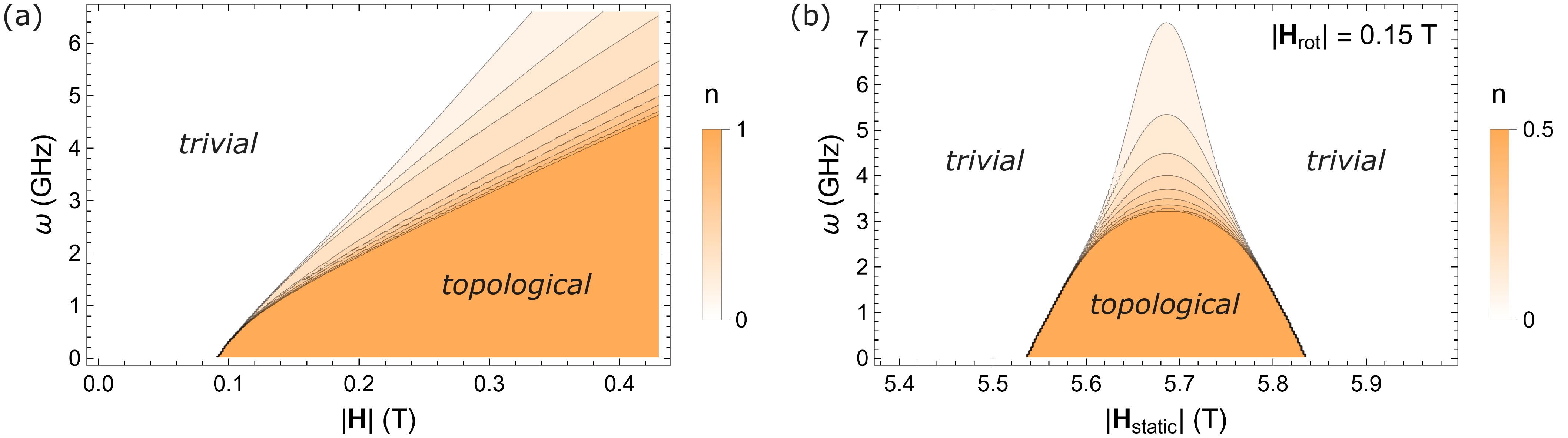}
	\caption{\label{fig:PhaseD}Dynamical phase diagram for bimeron pumping, defined in terms of $n$, the average bimeron displacement along $x$ per field rotation, in units of the spiral period. The values of $n$ are obtained by numerically solving the collective-coordinate equations of motion, using the parameters in Appendix~B and in-plane anisotropy $K_y = 0.01\;\mathrm{meV}$. (a) Ferromagnetic $J_\perp$ case: $\lvert{\bf H}\rvert$ is the amplitude of the rotating field and $\omega$ its angular frequency. The topological region ($n = w = 1$) corresponds to bimeron pumping by one spiral period for each field rotation, whereas in the trivial region ($n = 0$) the bimeron does not move. Between these two regimes, $n$ takes non-integer values, indicating the onset of non-adiabatic dynamics. (b) Antiferromagnetic $J_\perp$ case (spin $S = 1$), where the magnetic field has a rotating component with fixed amplitude $\lvert{\bf H}_\mathrm{rot}\rvert = 0.15 \;\mathrm{T}$ and a static component ${\bf H}_\mathrm{static}$ along $x$. The bell-shaped topological region is centered at the static field ${\bf H}_\mathrm{static}^*$ that compensates the in-plane anisotropy. Here, the bimeron is pumped by half a spiral period for each field rotation, corresponding to $n = w_\mathrm{AF}/2 = 1/2$.}
\end{figure}

To analyze how in-plane anisotropy influences the transition between adiabatic and non-adiabatic regimes, we define $n$ as the average bimeron displacement along $x$ per field rotation, expressed in units of the spiral period. In the adiabatic regime, this normalized displacement takes integer values and coincides with the winding number $w$, whereas in the non-adiabatic regime it can be non-integer. We use $n$ to construct the dynamical phase diagram in Fig.~\ref{fig:PhaseD}(a), which illustrates the transition from the topological regime ($n = w = 1$) to the non-adiabatic regime, where $n$ becomes non-integer and eventually drops to zero at higher frequencies. For $\lvert{\bf H}\rvert < \mathrm{H}^*$, the bimeron is pinned by the anisotropy potential \eqref{eq:HKy}, and $n = 0$ for any $\omega$. For $\lvert{\bf H}\rvert \gg \mathrm{H}^*$, the non-adiabatic transition occurs at $\omega = \omega^*$,  consistent with \eqref{eq:EqSteady}, and the boundary of the topological region is a straight line in the $(\lvert{\bf H}\rvert,\, \omega)$ plane. For $\lvert{\bf H}\rvert \gtrsim \mathrm{H}^*$, the transition occurs at lower frequencies, and the region with non-integer $n$ is sharper.

An oscillating field, such as ${\bf H}_y(t) = H_y \left(0,\sin(\omega t),0\right)$ or ${\bf E}_z(t) = E_z \left(0,0,\sin(\omega t)\right)$, cannot realize adiabatic pumping, since the effective potential for the bimeron periodically vanishes. A tilted in-plane anisotropy prevents this flattening and enables adiabatic pumping. Defining the tilt angle with respect to the spiral wave vector as $\theta_K \in (-\pi/2,\pi/2)$, we get the following expression for the bimeron energy at ${\bf H} = 0$,
\begin{equation}
\label{eq:HK45}
     H = H_0 - C_E \frac{\gamma q_\mathrm{top}}{Q} E_z \sin(\omega t) \sin(Q\Delta\bar{x}) -C_{K_y} K \frac{\lambda}{Q}\cos^2{\!(Q\Delta\bar{x}-\theta_K)},
\end{equation}
where $K>0$ is the tilted in-plane anisotropy. Similar to the case above, topological pumping occurs only when the oscillating electric field exceeds a critical amplitude $E_z^* \approx C_{K_y} K \lambda / C_E \gamma$. The direction of the pumping depends on the tilt angle $\theta_K$. For $-\pi/2< \theta_K < 0$, the bimeron moves along the spiral wave vector, whereas for $0 < \theta_K < \pi/2$, it is pumped in the opposite direction.


\section{Néel bimeron pumping}
Thus far, we have considered materials with a ferromagnetic exchange $J_\perp$ between neighboring spiral chains. When $J_\perp$ is antiferromagnetic, the spin texture can be described by the Néel vector ${\bf n}({\bf r}) = ({\bf m}_1({\bf r}) - {\bf m}_2({\bf r}))/2$ and the total magnetization ${\bf m}({\bf r}) = {\bf m}_1({\bf r}) + {\bf m}_2({\bf r})$, where ${\bf m}_1({\bf r})$ and ${\bf m}_2({\bf r})$ are the magnetization fields in the two antiferromagnetic sublattices. Integrating out ${\bf m}({\bf r})$, the Lagrangian and Rayleigh dissipation functional become \cite{Dasgupta21, Kim14}
\begin{equation}
\label{eq:LeRAF}
L_\mathrm{AF} = \int{dxdy \: \frac{1}{8 J_\perp}} (\dot{{\bf n}}^2 - 2 \mu{\bf H}\cdot{\bf n}\times\dot{\bf n}) - H_\mathrm{AF}, \qquad\qquad R_\mathrm{AF} = \frac{\alpha}{2} \int{dxdy \: \dot{{\bf n}}^2},
\end{equation}
where ${\bf H}$ is the applied magnetic field, and the Ginzburg-Landau free energy for ${\bf n}({\bf r})$ reads
\begin{equation}
\label{eq:HAF}
     H_\mathrm{AF} = \int{dxdy \left(J_x \left[-Q^2 (\partial_x {\bf n})^2 + \frac{1}{2} (\partial_{x}^2 {\bf n})^2\right] + \frac{1}{2} J_\perp (\partial_y {\bf n})^2 + K_z n_z^2 + \frac{\mu^2}{8 J_\perp}({\bf n}\cdot{\bf H})^2 \right)}.
\end{equation}
At ${\bf H} = 0$, $H_\mathrm{AF}$ is formally identical to the Ginzburg-Landau free energy for a ferromagnetic $J_\perp$, \eqref{eq:H0}. Hence, the structure of a ${\bf n}$-bimeron is the same as discussed for a ${\bf m}$-bimeron. However, the interaction with an applied magnetic field resembles quadratic anisotropy with a hard axis along ${\bf H}$. Thus, a static magnetic field ${\bf H}_\mathrm{static}^* = (2 \sqrt{2 J_\perp \lvert K_y \rvert}/\mu)\,\hat{e}$ can be applied to compensate for the effects of $K_y>0$, with $\hat{e}=\hat{x}$, and $K_y<0$, with $\hat{e}=\hat{y}$.

In the absence of in-plane anisotropy $K_y$, the bimeron energy takes the form
\begin{equation}
\label{eq:HAFBimeron}
     H_\mathrm{AF} = H_0^\mathrm{(AF)} - C_M^{\mathrm{(AF)}} \frac{\mu^2}{8 J_\perp}\frac{\lambda}{Q} \left[{\bf H} \cdot \left(\cos(Q\Delta\bar{x}), \sin(Q\Delta\bar{x}), 0\right)\right]^2 \!,
\end{equation}
where the dimensionless shape factor $C_M^{\mathrm{(AF)}}$ coincides with $C_{K_y} = 2\pi\sqrt{2/3}$.

Consider now the dynamics driven by the rotating magnetic field. For $\omega \ll \mu \lvert{\bf H}\rvert \ll \alpha J_\perp$, the first terms in $L_\mathrm{AF}$ are negligible, and the equations of motion for the bimeron become
\begin{equation}
\label{eq:EqFieldsAF}
    \Delta\dot{\bar{x}} = -\frac{\mu^2 \lvert{\bf H}\rvert^2 C_M^{\mathrm{(AF)}}}{8\alpha J_\perp C_{\Delta\bar{x}}Q} \sin(2\left(Q\Delta\bar{x} - \omega t\right)), \qquad\qquad \frac{\dot{\bar{y}}}{\Delta\dot{\bar{x}}} = \mathcal{O}\!\left(\! \frac{\mu \lvert{\bf H}\rvert}{\alpha J_\perp} \!\right)\! .
\end{equation}
Using the Ansatz \eqref{eq:Ansatz}, we can estimate the dimensionless shape factor $C_{\Delta\bar{x}} = 10\pi\sqrt{2/3}$. In contrast to the ferromagnetic case (see \eqref{eq:EqFields2}), the $y$-component of the bimeron velocity $\dot{\bar{y}}$ is negligible compared to the $x$-component $\Delta\dot{\bar{x}}$. Indeed, there is no gyrotropic force in the Néel vector dynamics. Nevertheless, bimeron motion still presents the two regimes described by \eqref{eq:2Reg}, but now the critical rotation frequency is
\begin{equation}
\label{eq:CritAF}
    \omega^*_\mathrm{AF} = \frac{\mu^2 \lvert{\bf H}\rvert^2 C_M^{\mathrm{(AF)}}}{8\alpha J_\perp C_{\Delta\bar{x}}}.
\end{equation}
Since $\omega^*_\mathrm{AF}$ is due only to damping, it is inversely proportional to $\alpha$. In contrast, the gyrotropic force dominates in the ferromagnetic case, and $\omega^*$ is directly proportional to $\alpha$, \eqref{eq:EqSteady}.

Alternatively, we consider the case where the in-plane anisotropy is nonzero but compensated by the static field ${\bf H}_\mathrm{static}^*$ discussed above. If the amplitude of the rotating field $\lvert{\bf H}_\mathrm{rot}\rvert$ is much smaller than $\lvert{\bf H}_\mathrm{static}^*\rvert$, the equation of motion for $\Delta{\bar{x}}$ takes the form
\begin{equation}
\label{eq:CompKy}
    \Delta\dot{\bar{x}} = \frac{\omega^*_\mathrm{AF2}}{2Q} \cos(2Q\Delta\bar{x} - \omega t + \phi_{K_y}), \qquad\qquad \omega^*_\mathrm{AF2} = \frac{\mu \lvert{\bf H}_\mathrm{rot}\rvert C_M^{\mathrm{(AF)}}}{\alpha C_{\Delta\bar{x}}} \sqrt{\frac{2 \lvert K_y \rvert}{J_\perp}},
\end{equation}
where $\omega^*_\mathrm{AF2}$ is the critical rotation frequency, the static magnetic field is expressed in terms of $K_y$, and the phase factor $\phi_{K_y}$ equals $0$ for $K_y<0$ and $\pi/2$ for $K_y>0$. For $\omega \leq \omega^*_\mathrm{AF2}$, the bimeron adiabatically follows a minimum of the sliding potential, leading to the steady-state solution $\Delta\bar{x} = (\omega/2Q)t + \Delta x_0$. Consequently, the bimeron moves by half a spiral period per full rotation of the field, in contrast to the previous cases where the displacement is one full period. According to the definition of winding number introduced in the ferromagnetic case, such a pumping cycle would correspond to a fractional winding number, $w=1/2$. However, in the antiferromagnetic case, the potential induced by the magnetic field has half the period of the spiral. Redefining the winding number $w_\mathrm{AF}$ with respect to this halved periodicity, the case of \eqref{eq:CompKy} corresponds to $w_\mathrm{AF}=1$, while the case of \eqref{eq:EqFieldsAF} corresponds to $w_\mathrm{AF}=2$. Following the same procedure as in the previous section, we construct the dynamical phase diagram in the $(\lvert{\bf H}_\mathrm{static}\rvert,\, \omega)$ plane, see Fig.~\ref{fig:PhaseD}(b). The topological region, now characterized by $n = w_\mathrm{AF}/2 = 1/2$, takes the shape of a bell centered at $\lvert{\bf H}_\mathrm{static}^*\rvert$, with a half-width set by the rotating-field amplitude.

The $\mu{\bf H}_\mathrm{static}^*\cdot{\bf n}\times\dot{\bf n}$ term of the Lagrangian \eqref{eq:LeRAF} gives rise to the transverse velocity
\begin{equation}
\label{eq:CompKyVy}
    \dot{\bar{y}} = - C_{\mathrm{H}_s} \frac{Q\lambda}{\alpha} \sqrt{\frac{\lvert K_y \rvert}{J_\perp}} \sin(Q\Delta\bar{x} + \phi_{K_y}) \, \Delta\dot{\bar{x}} + \mathcal{O}\!\left(\! \frac{\mu \lvert{\bf H}_\mathrm{rot}\rvert}{\alpha J_\perp} \!\right)\!,
\end{equation}
where ${\bf H}_\mathrm{static}^*$ is expressed in terms of the in-plane anisotropy $K_y$. Using the Ansatz \eqref{eq:Ansatz}, we estimate $C_{\mathrm{H}_s} = \sqrt{3}\,\pi/8$. For $\omega \leq \omega^*_\mathrm{AF2}$, $\Delta\bar{x}$ satisfies the steady-state solution of \eqref{eq:CompKy} and, therefore, $\dot{\bar{y}}$ oscillates with twice the period of the rotating field. After two full rotations of the field, the bimeron is pumped by one spiral period along $x$, while the net displacement along $y$ equals zero. Thus, the overall pumping occurs only along the spiral wave vector, as in the absence of in-plane anisotropy and static magnetic field.

\section{Conclusions}
We have shown that centrosymmetric spiral magnets host bimerons, localized topological spin textures similar to skyrmions, which carry magnetic and ferroelectric dipole moments. Remarkably, these dipole moments depend on the bimeron position with respect to the spiral background. Therefore, an electromagnetic field acts directly on the translational modes of this topological defect, in contrast to bimerons and skyrmions in ferromagnets. Specifically, a slowly rotating magnetic field pumps bimerons along the spiral wave vector by one period per field rotation. This topological pumping is characterized by the integer winding number $w=1$ and remains robust against perturbations, such as in-plane anisotropy, as long as the magnetic field exceeds a critical amplitude. However, it breaks down when the field rotates faster than a critical frequency because of gyrotropic effects, i.e. Berry phase contributions, or damping.
If adjacent spin-spiral chains interact ferromagnetically, the bimeron also moves perpendicular to the spiral wave vector with a velocity proportional to its topological charge. In contrast, topological charge contributions cancel between two antiferromagnetically interacting spin chains, and the bimeron only moves along the spiral wave vector.
In the presence of strong in-plane anisotropy, the rotating field has to overcome a significant energy barrier in order to pump the bimeron. In the antiferromagnetic case, this barrier can be suppressed by applying an additional static magnetic field. As a result, the bimeron can be pumped by a low-amplitude rotating field, but the displacement per field rotation is reduced to half the spiral period. In spiral multiferroics, bimerons correspond to ferroelectric domains and can be nucleated by an applied electric field, then inflated into larger bubble domains detectable, e.g., by second harmonic generation \cite{Hoffmann11, Hoffmann13}. We expect these bubble domains to be pumped similarly to bimerons under the action of a rotating magnetic field.

The results rely only on the symmetries of the spiral background and naturally extend to three-dimensional systems, for which the bimeron texture repeats for every fixed $z$, forming a bimeron string. Therefore, bulk spiral multiferroics, such as TbMnO$_3$, MnWO$_4$, and CuO \cite{Kimura03, Cheong07, Tokura10}, represent promising platforms for the realization of bimeron topological pumping. In these materials, spins in adjacent spiral chains typically interact antiferromagnetically. In contrast, monolayer multiferroics present ferromagnetically coupled spiral chains, e.g. NiI$_2$, VI$_2$, and NiBr$_2$ \cite{Song22, Sodequist23}. Hence, both the dynamical cases can be accessed. Nevertheless, the antiferromagnetic case is probably more relevant for racetrack memory applications, as the bimeron moves along a fixed direction, the spiral wave vector. Thus, bulk spiral multiferroics realize a natural racetrack, where a $2\pi$ field rotation shifts skyrmionic textures by precisely one spiral period.


\section{Acknowledgments}
N.N. was supported by JSPS KAKENHI Grant Numbers 24H00197 and 24H02231. N.N. was supported by the RIKEN TRIP initiative.

\section{Appendix A: Collective coordinates approach}
The damping matrix, the gyrotropic matrix, and the conservative forces entering \eqref{eq:Thiele} take the form \cite{Tretiakov08, Clarke08}
\begin{eqnarray}
\label{eq:Coeff}
    \Gamma_{IJ} = \int{dxdy \: \frac{\partial {\bf m}}{\partial \xi_I} \cdot \frac{\partial {\bf m}}{\partial \xi_J}} , \qquad G_{IJ} = \int{dxdy \: {\bf m} \cdot \left[ \frac{\partial {\bf m}}{\partial \xi_I} \times \frac{\partial {\bf m}}{\partial \xi_J}\right]} , \qquad F_I = - \frac{\partial H}{\partial \xi_I} . 
\end{eqnarray}
The bimeron symmetries \eqref{eq:Sym1} and \eqref{eq:Sym2} result in $\Gamma_{\Delta\bar{x}\,\bar{y}} = \Gamma_{\phi\,\bar{y}} = G_{\Delta\bar{x}\,\phi} = 0$. Combining the definition of topological charge \eqref{eq:TopQ} with the relation between mode-derivatives and coordinate-derivatives, we get $G_{\Delta\bar{x}\,\bar{y}} = 4\pi q_\mathrm{top}$ and $G_{\phi\,\bar{y}} = 4\pi q_\mathrm{top}/Q$. Neglecting the bimeron contribution over the spiral one, we obtain $\Gamma_{\phi\,\phi} = L_xL_y$, where $L_xL_y$ is the area of the system. Thus, the damping matrix $\Gamma_{IJ}$ and the gyrotropic matrix $G_{IJ}$ take the form ($I,J = \Delta\bar{x},\,\bar{y},\,\phi$)
\begin{equation}
\label{eq:GammaG}
\begin{aligned}
    \Gamma_{IJ} = \begin{pmatrix} Q\lambda C_{\Delta\bar{x}} & 0 & \lambda C_{\Delta\bar{x}\,\phi} \\ 0 & C_{\bar{y}}/Q\lambda & 0 \\ \lambda C_{\Delta\bar{x}\,\phi} & 0 & L_xL_y \end{pmatrix}\!, \qquad G_{IJ} = \begin{pmatrix} 0 & 4\pi q_\mathrm{top} & 0 \\ -4\pi q_\mathrm{top} & 0 & -4\pi q_\mathrm{top}/Q \\ 0 & 4\pi q_\mathrm{top}/Q & 0 \end{pmatrix}\!,
\end{aligned}
\end{equation}
where $C_{\Delta\bar{x}}$, $C_{\bar{y}}$ and $C_{\Delta\bar{x}\,\phi}$ are dimensionless shape factors. Using the Ansatz \eqref{eq:Ansatz}, we can estimate such factors, obtaining $C_{\Delta\bar{x}} = 10\pi\sqrt{2/3}$, $C_{\bar{y}} = 4\pi\sqrt{2/3}$ and $C_{\Delta\bar{x}\,\phi} = 2\pi\sqrt{6}$. While $F_{\bar{y}} = F_\phi = 0$ due to translational symmetry, the conservative force acting on $\Delta\bar{x}$ is
\begin{equation}
\label{eq:F}
  F_{\Delta\bar{x}} = C_E \gamma q_\mathrm{top} E_z \cos(Q\Delta\bar{x}) + C_M \mu\lambda {\bf H} \cdot \left(\sin(Q\Delta\bar{x}), -\cos(Q\Delta\bar{x}), 0\right)\!.
\end{equation}
An electromagnetic field exerts a force on the bimeron that depends on its position relative to the spiral background $\Delta\bar{x}$. Neglecting second-order terms in $\alpha \ll 1$ and combining \eqref{eq:Thiele}, \eqref{eq:GammaG}, and \eqref{eq:F}, we get the equations of motion \eqref{eq:Phi} and \eqref{eq:EqFields2} for the bimeron under the action of a rotating magnetic field ${\bf H}(t) = \lvert{\bf H}\rvert \left(\cos(\omega t),\sin(\omega t),0\right)$.

\section{Appendix B: Atomistic spin dynamics simulations}
To support the analytical results, we perform various atomistic spin dynamics simulations using the UppASD code \cite{Skubic08}, which numerically solves the LLG equations, i.e. the equations of motion derived from \eqref{eq:LeR}, for a lattice model. Consequently, the numerical results do not involve the continuum approximation and the collective coordinates approach. Defining ${\bf m}_{\bf r}$ as the unit vector along the spin at the lattice site ${\bf r}$, the simplest lattice model described by \eqref{eq:H0} takes the form
\begin{equation}
\label{eq:Disc}
H = \sum_{\bf r} \left(J_1 \, {\bf m}_{\bf r} \cdot {\bf m}_{{\bf r} + \hat{x}} + J_2 \, {\bf m}_{\bf r} \cdot {\bf m}_{{\bf r} + 2\hat{x}} + J_\perp \, {\bf m}_{\bf r} \cdot {\bf m}_{{\bf r} + \hat{y}} + K_z \left({\bf m}_{\bf r} \cdot \hat{z}\right)^2 - \mu {\bf H} \cdot {\bf m}_{\bf r}\right)
\end{equation}
with competing nearest neighbor $J_1 < 0$ and next nearest neighbor $J_2 > 0$ exchange interactions in the $x$-direction, ferromagnetic $J_\perp < 0$ in the $y$-direction, hard-$z$ anisotropy $K_z > 0$, magnetic moment of a single spin $\mu$ and applied magnetic field ${\bf H}$. At ${\bf H} = 0$, the continuum limit of \eqref{eq:Disc} is equivalent to \eqref{eq:H0} with $Q = \arccos(-J_1/4 J_2)$ and $J_x = J_2 \sin^2\!Q/Q^2$.

We first relax the bimeron Ansatz \eqref{eq:Ansatz} with the following parameters: $J_1 = -1.4 \; \mathrm{mRy}$, $J_2 = 0.38 \; \mathrm{mRy}$, $J_\perp = -0.4 \; \mathrm{mRy}$, $K_z = 0.008 \; \mathrm{mRy}$, $\mu=\mu_{\mathrm{B}}$ and $\alpha = 0.1$, where $\alpha$ denotes the Gilbert damping. The relaxed spin texture is qualitatively very similar to the Ansatz and continues to obey the symmetries defined in \eqref{eq:Sym1} and \eqref{eq:Sym2}. We then apply a rotating magnetic field ${\bf H}$ in the $xy$-plane with $\lvert{\bf H}\rvert = 0.1\;\mathrm{T}$ and angular frequency $\omega \in \left[0.25,5\right]\!\; \mathrm{GHz}$. The results, shown in Fig.~\ref{fig:Breakdown}, are in excellent agreement with the analytical solution, \eqref{eq:2Reg}, once the spin $S = \hbar/2$ and the lattice constant $a = 0.5\;\mathrm{nm}$ are restored. In accordance with \eqref{eq:EqFields2}, the bimeron also moves in the $y$-direction. The ratio between the velocities along $y$ and along $x$ does not depend on $\omega$ and is approximately $-21.6$ with this choice of parameters. We perform a similar set of simulations with an antiferromagnetic $J_\perp>0$ in the $y$-direction. The motion in the $x$-direction is equivalent to the previous case, with the critical frequency given by \eqref{eq:CritAF}. However, the bimeron no longer moves along $y$, as predicted by \eqref{eq:EqFieldsAF}.


%

\end{document}